\documentclass[%
 reprint,
 amsmath,amssymb,
aps,
prx,
floatfix,
]{revtex4-1}

\usepackage{graphicx}
\usepackage{dcolumn}
\usepackage{bm,xcolor}
\graphicspath{ {./figures/} }

\usepackage{amsmath,amssymb}
\begin{document}

\preprint{APS/123-QED}

\title{Relativistic corrections to the Di{\'o}si-Penrose model}

\author{Luis A. Poveda}
\email{poveda@cefetmg.br}
\affiliation{Departemento de F\'isica, Centro Federal de Educa\c{c}\~ao Tecnol\'ogica de Minas Gerais\\
Amazonas 5253, 30421-169 Belo Horizonte, MG, Brasil
}%

\author{Luis Grave de Peralta}
\email{Luis.Grave-de-Peralta@ttu.edu}
\affiliation{Department of Physics and Astronomy, Texas Tech University, Lubbock, TX 79409, USA
}%

\author{Arquimedes Ruiz-Columbi\'e}
\email{Arquimedes.Ruiz-Columbie@ttu.edu}
\affiliation{Wind Energy Program, Texas Tech University, Lubbock, TX 79409, USA
}%



\date{\today}

\begin{abstract}
The Di{\'o}si-Penrose model is explored in a relativistic context. Relativistic effects were considered within a recently proposed Grave de Peralta approach [L. Grave de Peralta, {\em Results Phys.} {\bf 18} (2020) 103318], which parametrize the Schr{\"o}dinger-like hamiltonian so as to impose that the average kinetic energy of the system coincide with its relativistic kinetic energy. As a case of study, the method is applied to a particle in a box with good results. In the Di{\'o}si-Penrose model we observed that the width of a quantum matter field confined by its own gravitational field [L. Di{\'o}si, {\em Phys. Lett}. {\bf 105A} (1984) 199], sharply drop to zero for a mass of the order of the Planck mass, indicating a breakdown of the model at the Planck scale.

\end{abstract}

\pacs{Valid PACS appear here}
\maketitle


\section{\label{sec:Intro}Introduction}

The Schr{\"o}dinger-Newton equation~\cite{Diosi1984,Moroz1998,Harrison2003,Robertshaw2006,Adler2007,Bahrami2014} is a model which describe the time evolution of a Schr{\"o}dinger quantum field coupled to a Newtonian gravitational field. This have been called the Di{\'o}si-Penrose collapse model~\cite{Diosi1984,Penrose1998,Penrose2014} aimed to elucidate the role of gravity on quantum state reduction~\cite{Karolyhazy1966}. For a single particle of mass $m$ the equation is written as
\begin{equation}
i\hbar \frac{\partial\Psi}{\partial t} =\left[-\frac{\hbar^2\nabla^2}{2m}-{Gm^2}\int\frac{~|\Psi|^2d\mathbf{x'}}{|\mathbf{x}-\mathbf{x'}|}\right]\Psi,
\label{Eq01}
\end{equation}
where the second term in the hamiltonian is the self gravitational potential due to the mass distribution $m|\Psi|^2$. 

As point out~\cite{Bahrami2014}, Eq.~(\ref{Eq01}) represents the weak gravitational field non-relativistic limit of a fundamentally semi-classical theory of gravity~\cite{Moller,Rosenfeld1963}. That is, according to~(\ref{Eq01}) gravity is a classical field even at the fundamental level, to which a quantum matter field is somehow coupled. As shown by Di{\'o}si in an early paper~\cite{Diosi1984}, a stationary solution of Eq.~(\ref{Eq01}) for a particle of mass $m$ has a width of the order
\begin{equation}
a_0=\frac{\hbar^2}{Gm^3},
\label{Eq02}
\end{equation}
representing the length at which the diffusion of the matter field is counteracted by its gravitational self force~\cite{Diosi1984}. 

Eq.~(\ref{Eq02}) can be written as
\begin{equation}
a_0=l_P\left(\frac{m_P}{m}\right)^3,
\label{Eq02a}
\end{equation}
where $l_P=\sqrt{\hbar G/c^3}$ and $m_P=\sqrt{\hbar c/G}$ are the Planck length and mass, respectively. Note that at $m=m_P$, $a_0$ match the Planck length, further implying that the mass distribution $m|\Psi(\mathbf{x},t)|^2$ occupies a region of the order of its Schwarzschild radius $2Gm/c^2$. Therefore, the Eq.~(\ref{Eq01}) should be valid only far from the Planck scale.

In this work we explore qualitatively the coupling between a stationary relativistic scalar matter field and its classical self gravitational field, through a quasi-relativistic Schr{\"o}dinger-like model. This is done within the recently proposed Grave de Peralta approach~\cite{Grave2020:196,Grave2020:103318,Grave2020:788,Grave2020:14925,Grave2020:065404}, which include relativistic effects by a suitable parametrization of the Sch{\"o}dinger hamiltonian. Our analysis suggest that, in the relativistic regime, a characteristic length analogous to Eq.~(\ref{Eq02}) can be indentified, involving the (reduced) Compton wavelength of the particle, $\lambdabar_C=\hbar/mc$. This relativistic Di{\'o}si length start to deviates from $a_0$ when the mass of the particle approach a mass of order of the Planck mass and right after it drop to zero sharply. 

The paper is organized as follow. In section~\ref{sec:grave}, the Grave de Peralta approach to relativity is briefly summarized; in section~\ref{sec:box} the method is apply to the quantum problem of a particle in the infinite one-dimensional well; a analysis of Di{\'o}si-Penrose model is presented in section~\ref{sec:diosi}; some conclusions are given in section~\ref{sec:concluc}.

\section{\label{sec:grave}Grave de Peralta equation}

A relativistic quantum theory for a spin-$0$ article of mass $m$ is formally obtained from first quantization of the energy momentum relation
\begin{equation} 
E=\sqrt{m^2c^4+c^2p^2}
\label{Eq03}
\end{equation} 
a route which lead to a cumbersome square root operator (SRO). 

The Grave de Peralta (GdeP) approach take advantage of the great similitude between the relativistic kinetic energy~\cite{Christodeulides}
\begin{equation} 
\frac{{p}^2}{\left(1 + \gamma_v \right)m},
\label{Eq03a}
\end{equation} 
and its non-relativistic counterpart $p^2/2m$, to propose a quasi-relativistic Schr{\"o}dinger-like equation as follows.

In Eq.~(\ref{Eq03a}), $\gamma_v=\left({1-v^2/c^2}\right)^{-1/2}$ is the Lorentz factor, which in term of the momentum and mass of the particle adopt the form, 
\begin{equation} 
\gamma= \sqrt{1+\frac{{p}^2}{m^2 c^2}},
\label{Eq03b}
\end{equation}
such that the first quantization of Eq.~(\ref{Eq03a}) involve the SRO,   
\begin{equation} 
\hat{\gamma}= \sqrt{1+\frac{\hat{p}^2}{m^2 c^2}},
\label{Eq03d}
\end{equation}

In order to avoid the SRO, Grave de Peralta defined $\gamma$ as a constant parameter, which value tends to $1$ in the non-relativistic regime ($v\ll c$ or $p\ll mc$), and construct the following quasi-relativistic kinetic energy operator~\cite{Grave2020:196,Grave2020:103318,Grave2020:788,Grave2020:14925,Grave2020:065404},
\begin{equation} 
\frac{{\hat p}^2}{\left(1 + \gamma \right)m},
\label{Eq03e}
\end{equation} 
where $\hat{p}\equiv -i\hbar\nabla$. After replacing the kinetic energy operator ${\hat p}^2/2m$ in the stationary Schr{\"o}dinger equation by the operator~(\ref{Eq03e}), the following stationary equation is obtained,
\begin{equation} 
\left[\frac{-\hbar^2\nabla^2}{(1+\gamma)m}+\hat{V} \right]\psi=E\psi,
\label{Eq04}
\end{equation} 

We call Eq.~(\ref{Eq04}) the Grave de Peralta (GdeP) equation and its relativistic character rely on the requirement that the parameter $\gamma$ enforce that the average of the operator~(\ref{Eq03e}) in a certain stationary state, $\chi$, coincide with the relativistic kinetic energy of the particle, that is, 
\begin{equation} 
\bigg \langle\chi \left|\frac{\hat{p}^2}{(1+\gamma)m} \right|\chi \bigg\rangle=mc^2(\gamma-1),
\label{Eq05}
\end{equation} 

After solving for $\gamma$ we obtain
\begin{equation} 
\gamma=\sqrt{1+\frac{\langle\chi \left|\hat{p}^2\right|\chi \rangle}{m^2c^2}},
\label{Eq06}
\end{equation} 
or
\begin{equation} 
\gamma=\sqrt{1+\frac{2}{mc^2}\langle\chi |\hat{H}-\hat{V}|\chi \rangle},
\label{Eq07}
\end{equation} 
where $\hat{H}=\hat{p}^2/2m+\hat{V}$. 

Then by choosing $\chi$ as an {\em eigen}vector of $\hat{H}$ with {\em eigen}value $\mathcal{E}$, the Eq.~(\ref{Eq07}) adopt the form,
\begin{equation} 
\gamma=\sqrt{1+\frac{2}{mc^2}\left(\mathcal{E}-\langle\chi |\hat{V}|\chi \rangle\right)},
\label{Eq08}
\end{equation} 

This method have been applied to well known quantum mechanical problems with good results~\cite{Grave2020:196,Grave2020:103318,Grave2020:788,Grave2020:14925,Grave2020:065404}, with the advantage that Eq.~(\ref{Eq04}) can be solved using common techniques of non-relativistic quantum mechanics. In the next section we show that the GdeP equation may include relativistic corrections in a reliable way, for a particle in a one-dimensional box of size $L$~\cite{Grave2020:103318}.

\section{\label{sec:box}Relativistic particle in a box}

In this case $\hat{V}=0$ inside the well ($0\le x\le L$), hence the relativistic energy is obtained directly from Eq.~(\ref{Eq04}),
\begin{equation}
E_n=\frac{2}{1+\gamma_n}\mathcal{E}_n,
\label{Eq12}
\end{equation}
where
\begin{equation}
\mathcal{E}_n=\frac{\hbar^2 n^2\pi^2}{2mL^2},
\label{Eq13}
\end{equation}
is the non-relativistic energy for the $n$-th level~\cite{Griffiths}, and for the $n$-th {\em eigen}state
\begin{equation} 
\gamma_n=\sqrt{1+\frac{\hbar^2n^2\pi^2}{m^2c^2L^2}}.
\label{Eq14}
\end{equation}

Substituting Eqs.~(\ref{Eq13}) and~(\ref{Eq14}) in Eq.~(\ref{Eq12}), the expression for the relativistic energy of the $n$-th level, within the Grave de Peralta approach will be,
\begin{equation}
E_n=\frac{\hbar^2 n^2 \pi^2 }{\left[ 1+\sqrt{ 1+\frac{\hbar^2n^2\pi^2}{m^2c^2L^2}}\right]m L^2},
\label{Eq15}
\end{equation}
which coincide with the expression obtained previously~\cite{Grave2020:103318}.

The above equation can be written in the following form
\begin{equation}
E_n=mc^2\sqrt{ 1+\frac{p_n^2}{m^2c^2}}-mc^2,
\label{Eq16}
\end{equation}
where $p_n=\hbar n\pi/L$ and, clearly, it is the relativistic kinetic energy involving the non-relativistic momentum {\em eigen}values. 

By expanding Eq.~(\ref{Eq16}) in power of $p_n^2$ we obtain
\begin{equation}
E_n=\frac{\hbar^2 n^2\pi^2 }{2mL^2}-\frac{\hbar^4 n^4\pi^4} {8m^3 c^2 L^4}+\mathcal{O}(L^{-6}),
\label{Eq17}
\end{equation}
where the first order coincides with the non-relativistic energy, Eq.~(\ref{Eq13}), and the second order gives the mass-velocity relativistic correction to the kinetic energy.
\begin{figure}
\centering
        \includegraphics[scale=1, width=8.6cm]{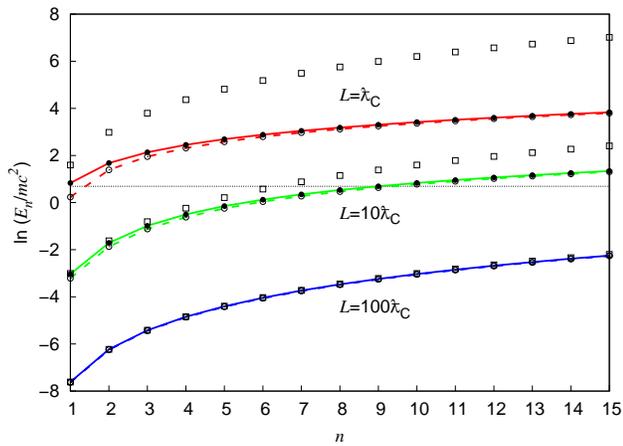}
        \caption{Comparison of the present results (solid dots, solid lines) from Eq.~(\ref{Eq13}) and the reported in ref.~\cite{Alberto1996} (empty dots, dashed lines), for different sizes of the well. In square dots the corresponding non-relativistic energies. The thin line indicates the energy threshold $E_n=2mc^2$.}
        \label{fig:1}     
\end{figure}

A comparison of the energies computed with Eq.~(\ref{Eq15}) and those reported in reference~\cite{Alberto1996}, is given in Figure~\ref{fig:1}, for different sizes of the well. Apart from some disagreements for lower values of $n$ above the threshold $E_n=2mc^2$, the present model shows a very good agreement with most of the reported data and describe the tendency when the energy increase and the size of the well decreases.

In general for a particle confined in a region of size $a$, the parameter $\gamma$ of Eq.~(\ref{Eq06}) can be evaluated as
\begin{equation}
\gamma\approx\sqrt{1+\left(\frac{\lambdabar_C}{a}\right)^2},
\label{Eq18}
\end{equation}
where $\lambdabar_C$ is the reduced Compton wavelength.

\begin{figure}
\centering
        \includegraphics[scale=1, width=8.6cm]{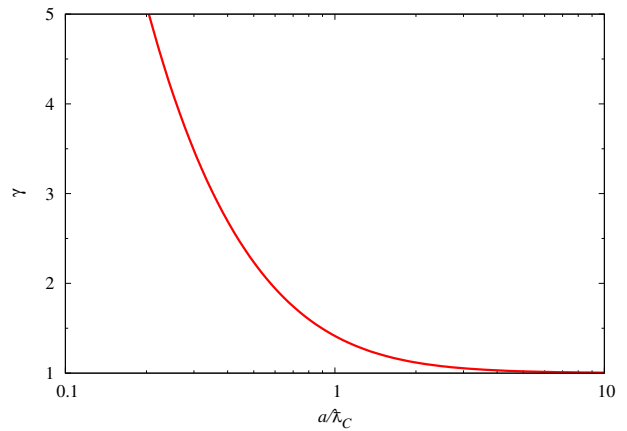}
        \caption{A plot of Eq.~(\ref{Eq18}). See text for details.}
        \label{fig:2}
\end{figure}

Figure~\ref{fig:2} show the behavior of $\gamma$ as a function of the particle seize in units of $\lambdabar_C$. From the figure, the parameter appreciable deviates from $1$ when the localization region of the particle approach the Compton wavelength. Moreover, when the particle extent to a spatial region several orders larger than $\lambdabar_C$, the relativistic effects becomes negligible. It is worth noting that for a particle with a mass close to the Planck mass, the relativistic effects start to become relevant when the particle is localized in a region of the order of the Planck length, which coincide with $\lambdabar_C$ for $m=m_P$.  

\section{\label{sec:diosi}The relativistic Di{\'o}si length}
The Eq.~(\ref{Eq01}) was first considered by Di{\'o}si as a model for the suppression of quantum behavior of a macroscopic object by its own gravitational field~\cite{Diosi1984}. After proving the existence of a stationary ground state solution, Di{\'o}si evaluated the width of the corresponding wave function, $\psi$, as the length at which the driven force, due to the diffusive kinetic energy term, is balanced by the self gravitational force of the mass distribution $m|\psi|^2$~\cite{Diosi1984}. Thus, by estimating the total energy as,
\begin{equation}
E\approx \frac{\hbar^2}{2m a_0^2}-\frac{Gm^2}{a_0}.
\label{Eq19}
\end{equation}
the length which minimize this expresion is given by Eq.~(\ref{Eq02}).

Now let us qualitatively evaluate the total energy for a relativistic quantum matter field of size $a$ in its own gravitational self potential as,  
\begin{equation}
E\approx \frac{\hbar^2}{(1+\gamma)m a^2}-\frac{Gm^2}{a},
\label{Eq20}
\end{equation}
where $\gamma$ is given by Eq.~(\ref{Eq18}).

It is straightforward to show that the value of $a$ for which Eq.~(\ref{Eq20}) has a minimum is
\begin{equation}
a=a_0\sqrt{1-\left(\frac{\lambdabar_C}{a_0}\right)^2},
\label{Eq21}
\end{equation}
or written in term of the mass of the particle
\begin{equation}
a=a_0\sqrt{1-\left(\frac{m}{m_P}\right)^4}.
\label{Eq22}
\end{equation}

\begin{figure}
\centering
        \includegraphics[scale=1, width=8.6cm]{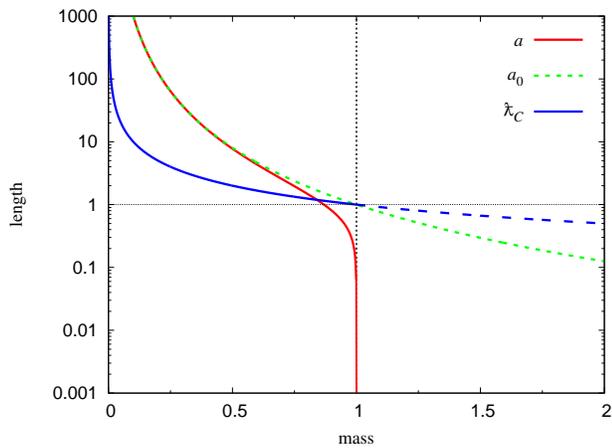}
        \caption{Length scales in units of the Planck length as a function of the mass in units of the Planck mass.}
        \label{fig:3}     
\end{figure}

Figure~\ref{fig:3} show the behavior of the different length scales in units of the Planck length as a function of the mass of the particle in units of the Planck mass. From the figure it is evident that close to the Planck mass the relativistic corrected Di{\'o}si length start to depart from $a_0$ and sharply drop to zero, then becoming undefined for larger vales. This behavior clearly indicate a breakdown of Eq.~(\ref{Eq01}) when a quantum matter field is localized, by its gravitational self attraction, below the Compton wavelength. Moreover, the Eq.~(\ref{Eq22}) suggest an interesting connection between the Planck mass and the reduction of the quantum wave packet induced by gravity, for which the characteristic length approach the Planck scale.

It is worth noting that Eq.~(\ref{Eq20}) can be written as,
\begin{equation}
E\approx mc^2\left[\sqrt{1+\left(\frac{\lambdabar_C}{a}\right)^2}-1\right]-\frac{Gm^2}{a},
\label{Eq23}
\end{equation}
and the first term in the right hand side is of the order of the relativistic kinetic energy.

\section{\label{sec:concluc}Conclusions}

In this work we applied the Grave de Peralta approach to include relativistic corrections to the quantum problem of a particle confined in the one-dimensional infinite potential well. The present results appears in good agreement with the values obtained by integration of the Dirac equation. Here we observed that a particle localized in a region of the order of its Compton wave length is strongly relativistic. The major advantage of this method rely on the possibility of include relativistic corrections by solving a Schr{\"o}dinger-like equation.

The method was then applied to qualitatively evaluate the width of a stationary relativistic quantum matter field confined by its own classical non-relativistic gravitational field. Here we observed that when the mass of the particle approach a mass of the order of the Planck mass this model breakdown as the width of quantum field sharply drop to zero, becoming undefined for larger masses.    


\section*{\label{sec:acknw}Acknowledgement}

We thanks professor Pedro Alberto for clarifying some points of ref.~\cite{Alberto1996}.

\bibliographystyle{aipnum4-1}

\bibliography{bibliodiosi}

\end{document}